\documentclass[aps,twocolumn,pre,showpacs]{revtex4}
\usepackage{graphics}

\begin{document}

\title{The nature of slow dynamics in a minimal model of
frustration-limited domains} 
\author{Phillip L. Geissler}
\affiliation{Department of Chemistry, Massachusetts Institute of
Technology, Cambridge, MA 02139} 
\author{David R. Reichman}
\affiliation{Department of Chemistry and Chemical Biology, Harvard
University, Cambridge, MA 02138}

\begin{abstract}
We present simulation results for the dynamics of a schematic model
based on the frustration-limited domain picture of glass-forming
liquids.  These results are compared with approximate theoretical
predictions analogous to those commonly used for supercooled liquid
dynamics.  Although model relaxation times increase by several orders
of magnitude in a non-Arrhenius manner as a microphase separation
transition is approached, the slow relaxation is in many ways
dissimilar to that of a liquid.  In particular, structural relaxation
is nearly exponential in time at each wave vector, indicating that the
mode coupling effects dominating liquid relaxation are comparatively
weak within this model.  Relaxation properties of the model are
instead well reproduced by the simplest dynamical extension of a
static Hartree approximation.  This approach is qualitatively accurate
even for temperatures at which the mode coupling approximation
predicts loss of ergodicity.  These results suggest that the {\em
thermodynamically disordered} phase of such a minimal model poorly
caricatures the slow dynamics of a liquid near its glass transition.
\end{abstract}
\pacs{64.70.Pf, 64.60.Cn, 05.70.Fh}
\maketitle

\section{introduction}
Several microscopic scenarios have been proposed as underlying
mechanisms for dynamical arrest in supercooled liquids
\cite{goetze,wolynes,sethna,parisi,chandler}.  Because this dramatic
slowing down is accompanied by the onset of dynamical
heterogeneity\cite{ediger,ediger2001,weitz}, a promising candidate
model should account for the spontaneous segregation of rapidly
relaxing and slowly relaxing domains.  The theory of
frustration-limited domains has been developed with this condition in
mind\cite{fldt,sachdev}.  The basic units of this theory are
microscopic regions of low internal energy whose spatial extent is
limited by long-ranged interactions or constraints.  Nelson and
coworkers suggested that such domains form in simple ``atomic''
liquids (such as metallic glass-formers)\cite{nelson}.  For small
clusters of particles, icosahedral arrangements are energetically
preferred over closest-packed configurations representative of the
crystalline state\cite{nelson1989}.  It is thus argued that a
supercooled atomic liquid is rich in low-energy icosahedral clusters
which, for geometric reasons, cannot extend indefinitely.  In this
case, frustration is a consequence of the vanishing curvature of
Euclidean space.  One can imagine that the nature of local order and
source of frustration are somewhat different for other, more
complicated materials.

The frustration-limited domain theory has several attractive features.
By associating relaxation kinetics with the interfacial area of domain
walls, it predicts a crossover in the temperature dependence of
structural rearrangement times from Arrhenius to a super-Arrhenius
form.  Such a crossover is prominent in experiments with fragile
glass-formers\cite{nagel}.  Further, the theoretical scaling exponent
for the asymptotic super-Arrhenius temperature dependence appears to
be consistent with experimental data.  Finally, the notion of
frustration-limited domains resonates with the picture of
heterogeneous dynamics that has emerged from
experiments\cite{ediger,ediger2001,weitz} and
simulations\cite{harrowell,onuki,glotzer}.  While local structure
within low-energy clusters may be effectively frozen on a molecular
time scale, relaxation can be facile at the strained interfaces
between domains.  This argument is the essence of Stillinger's ``tear
and repair'' picture of shear flow in fragile
liquids\cite{stillinger}.

Although these dynamical predictions of frustration-limited domain
theory are suggestive, they are fundamentally thermodynamic in nature
and thus indirect.  In order to make such arguments precise, Grousson
{\em et al.} have recently focused on the explicit dynamics of minimal
models exhibiting frustration-limited domains\cite{grousson_etal}.
Specifically, they have simulated stochastic dynamics of several
classical spin models that pit short-ranged, ferromagnetic
interactions against long-ranged, antiferromagnetic interactions.  In
addition to confirming super-Arrhenius relaxation for these models at
low temperatures, they have demonstrated that the ``fragility'' of the
dynamics (i.e., the degree of deviation from Arrhenius form) varies
continuously with the relative strength of the long-ranged
frustration.  Quite recently, Grousson {\em et al.} have performed
dynamical mode coupling calculations for the same models, but no
direct comparisons to their earlier simulations were
made\cite{grousson_etal2}.  In this paper, we perform similar
calculations that are compared directly to numerical simulations.

Schmalian and coworkers have also argued that such a
Coulomb-frustrated ferromagnet should display essential features of
glassy dynamics\cite{schmalian,schmalian2,schmalian3}.  In their
analysis, it is a proliferation of metastable states that drives
vitrification.  With the aid of replica mean field theory, this
perspective predicts a scaling of fragility with frustration strength
that agrees well with the simulation results of
Ref.\cite{grousson_etal}.  But like the theory of frustration-limited
domains, this analysis is {\em thermodynamic} in nature, relying on an
assumed correspondence between particular subensembles of high free
energy and genuine dynamical bottlenecks.

This paper addresses the extent to which the slow relaxation in such
simplified models truly resembles that of molecular glass-forming
liquids.  For this purpose we investigate in detail the {\em dynamics}
of a model closely related to those studied in
Refs.\cite{grousson_etal} and \cite{schmalian}.  The model and
dynamical propagation rules we consider, which are free of
artificially quenched disorder and kinetic constraints, are described
in Sec. 2.  For several values of the frustration strength, we compare
the time dependence of spin correlations computed in simulations with
those predicted by approximate theoretical approaches.

Two self-consistent dynamical equations obtained from theory are
discussed in Sec. 3.  They correspond to resummations of an exact,
infinite diagrammatic series for time correlations.  The first
resummation, yielding exponential relaxation at each wave vector, is a
direct dynamical generalization of Brazovskii's static result for this
class of models\cite{brazovskii}.  Calculations based on this
straightforward approach agree remarkably well with simulation
results, described in Sec. 4, even for temperatures approaching a
thermodynamic transition to a fully ordered state.  The second
resummation is formally analogous to the idealized mode coupling
theory of liquids.  As such, it predicts loss of ergodicity at finite
temperature.  By contrast, we find no evidence of nonergodic behavior
or two-step relaxation for simulated disordered states of this model.

Surprisingly, then, the Hartree approach is the more accurate
approximation for slowly relaxing disordered states of the model
system.  Detailed comparison of simulation and theory confirms that
sluggishness indeed arises from de~Gennes narrowing (i.e., from
dramatic changes in static correlations), in contrast to complex
dynamical mechanisms such as mode coupling.  In this respect, the
minimal model we have studied does not capture important aspects of
supercooled liquid dynamics, specifically the intermediate time
plateau and long time stretching of dynamical correlators.
Implications of this result are discussed in Sec. 5, along with issues
related to fragility and local conservation of magnetization.  In
section 6 we conclude.

\section{Model}
We consider fluctuations of a field $\phi({\bf r})$
at position ${\bf r}$ in three dimensions, with
energy\cite{brazovskii,hohenberg,chakraborty} 
\begin{eqnarray}
&& \hspace{-5mm}
\beta {\cal H}[\phi({\bf r})] =
\nonumber \\
&& \int d{\bf r} \, \left[\phi({\bf r})
\left(\tau + k_0^{-2}(\nabla^2 + k_0^2)^2\right) \phi({\bf r}) +
{\lambda \over 4!}\phi^4({\bf r})\right].
\label{equ:hamiltonian}
\end{eqnarray}
Here, the energy scale $\beta^{-1}$ characterizes typical fluctuations
of a surrounding heat bath.  The physical meaning of the field $\phi$
may be somewhat abstract in the context of supercooled liquids, for
example representing the degree of a particular local packing
symmetry.  The application of Eq.~\ref{equ:hamiltonian} to diblock
copolymer melts is more intuitive\cite{leibler,kawasaki,binder}.  In
this case, $\phi$ represents the local excess number density of one
monomer type, and $\tau$ describes the preferential affinity of
monomers for others of the same type.  Because our interest in this
model is motivated by the work in Ref.\cite{grousson_etal}, we will
imagine that $\phi$ simply represents a coarse-grained, scalar spin
density.  Here, $\tau$ is a dimensionless temperature measuring the
distance from an underlying critical temperature when $k_0 = 0$.  For
any physical interpretation, the wave vector $k_0 \neq 0$
characterizes long-ranged order of a low-temperature, microphase
separated state.  The coefficient $\lambda$ multiplying the $\phi^4$
term in Eq.~\ref{equ:hamiltonian} will later be used to order terms in
perturbation series, although in calculations its numerical value will
be of order unity.

For wave vectors near $k_0$, the action in Eq.~\ref{equ:hamiltonian}
corresponds to that studied by Schmalian and Wolynes
\cite{schmalian,schmalian2,schmalian3} and (in a hard-spin lattice
version) by Grousson {\em et al.}\cite{grousson_etal}.  In their work,
frustration is explicit in competing interactions of square-gradient
($J \, |\nabla \phi({\bf r})|^2$) and Coulomb ($Q \, \phi({\bf r})
\phi({\bf r}')/|{\bf r}-{\bf r}'|$) forms.  The relative strengths of
these interactions determine the periodicity of the ground state, $k_0
\propto (Q/J)^{1/4}$, in which spin-up and spin-down domains alternate
in stripes or lamellae\cite{scaling_footnote}.  In the model defined
by Eq.~\ref{equ:hamiltonian}, this frustration is instead implicit in
nonzero $k_0$, but has the same physical effect.  Namely, homogeneous
domains are energetically favored at small length scales, while net
magnetization is effectively constrained to vanish at larger length
scales.  In the context of diblock copolymers, this effective
constraint reflects the stoichiometry imposed by polymer
connectivity\cite{leibler,kawasaki,binder}.

In two dimensions and higher, the presence of nonzero $k_0$ in
Eq.~\ref{equ:hamiltonian} has a subtle but profound effect on the
thermodynamics of the paramagnetic state.  Specifically, the large
entropy of fluctuations near $|{\bf k}|=k_0$ significantly reduces the
free energy of the disordered phase.  Within a Hartree approximation,
this contribution is sufficient to make the paramagnetic state stable
or metastable for all finite $\tau$.  As a consequence, the transition
to a phase with long-ranged order is first (rather than second) order
and occurs at a temperature $\tau_{\rm tr} < 0$.  This effect was
first recognized by Brazovskii\cite{brazovskii} and has been
summarized lucidly by Binder and Fredrickson\cite{binder}.  Its
qualitative features have been subsequently confirmed in experiments
with diblock copolymers\cite{fredrickson}.  Because, in this picture,
statistics of the disordered state are dominated by fluctuations near
$k_0$, we expect the model of Eq.~\ref{equ:hamiltonian} to belong to
the same universality class as those of Refs.\cite{grousson_etal} and
\cite{schmalian,schmalian2,schmalian3}.  Later, we will demonstrate
that slow dynamics of this state are dominated by the very same
fluctuations.

The action in Eq.~\ref{equ:hamiltonian} is not a true Hamiltonian, and
thus has no intrinsic dynamics.  We consider two commonly used
stochastic propagation rules which generate a canonical ensemble of
fluctuations consistent with Eq.~\ref{equ:hamiltonian}.  Attention
will be primarily focused on a simple Langevin equation,
\begin{equation}
{\partial \phi({\bf r}) \over {\partial t}} = -{\delta {\cal H}[\phi]
\over{\delta \phi({\bf r})}} +\eta({\bf r},t),
\label{equ:langevin}
\end{equation}
where $\eta({\bf r},t)$ is a random force whose statistics are
Gaussian, and
\begin{equation}
\langle \eta({\bf r},t) \eta({\bf r}',t') \rangle = 2 \beta^{-1}
\delta ({\bf r}-{\bf r}') \delta (t-t').
\label{equ:noise}
\end{equation}
In Eq.~\ref{equ:noise}, angled brackets denote an average over all
possible realizations of the random force.  The above equation of
motion, along with the energetics of Eq.~\ref{equ:hamiltonian} and the
statistics of Eq.~\ref{equ:noise}, has been studied previously, most
notably in the context of nucleation and nonequilibrium pattern
formation following a rapid quench to $\tau < \tau_{\rm
tr}$\cite{hohenberg,chakraborty,elder,bray}.  To our knowledge, the
detailed equilibrium dynamics of the paramagnetic phase very close to
the transition (i.e., $\tau \gtrsim \tau_{\rm tr}$) have not until now
been fully explored.

The dynamics generated by Eq.~\ref{equ:langevin} do not conserve the
field $\phi({\bf r},t)$\cite{bray}.  Because the slow relaxation of
supercooled liquids results in part from the conservation of
hydrodynamic densities, this feature of Eq.~\ref{equ:langevin} may be
somewhat troubling (particularly in the context of diblock copolymers,
in which the number density is clearly conserved).  For this reason,
we consider a second form of dynamics that conserves $\phi({\bf r},t)$
by construction.  Trajectories of this dynamics are chains of
microstates generated by a Metropolis Monte Carlo algorithm.  In
detail, a random displacement of the field $\phi({\bf r})$ is
attempted at discrete time steps, and is accepted with probability
\begin{equation}
P_{\rm acc} = \min\left[ 1,\exp{(-\beta\Delta{\cal H})} \right],
\label{equ:mc}
\end{equation}
where $\Delta {\cal H}$ is the change in energy produced by the
displacement.  Local conservation of the field is achieved by
restricting the choice of random displacements to those which do not
alter the local net magnetization.  Further details of this procedure
will be described in Sec. 4.  Simulation results presented in that
section demonstrate that the decay of spin correlations produced by
Eqs.~\ref{equ:langevin} and~\ref{equ:mc} are nearly identical, within
an arbitrary rescaling of time in the Monte Carlo chain of states.
The physical meaning of this fact will be discussed in Sec. 5.

\section{Theory}
In this section we discuss two approximations for relaxation of the
field $\phi({\bf r},t)$.  Specifically, we derive closed equations of
motion for the correlation function
\begin{equation}
C_{\bf k}(t-t')=\langle \phi_{{\bf k}}(t)\phi_{-{\bf k}}(t')\rangle,
\end{equation}
where Fourier components of the field are defined in the standard way:
\begin{equation}
\phi_{{\bf k}}(t) = \int d{\bf r} \, \phi({\bf r},t) e^{i{\bf
k}\cdot{\bf r}}.
\end{equation}
The first equation of motion is linear in $\phi_{{\bf k}}(t)$ and
resembles phenomenological theories of high temperature liquid state
dynamics, such as the wave vector dependent viscoelastic
theory\cite{hansen}.  The second is nonlinear and has the form of the
idealized mode-coupling approximation to the dynamics of density
fluctuations in supercooled liquids\cite{goetze}.  This equation
contains the feedback mechanism responsible for the interruption of
particle diffusion at intermediate time scales due to constraints
imposed by slowly reorganizing local environments (the ``cage''
effect).

\begin{figure}
\vspace*{1cm}
\centerline{\resizebox{6cm}{!}{\includegraphics{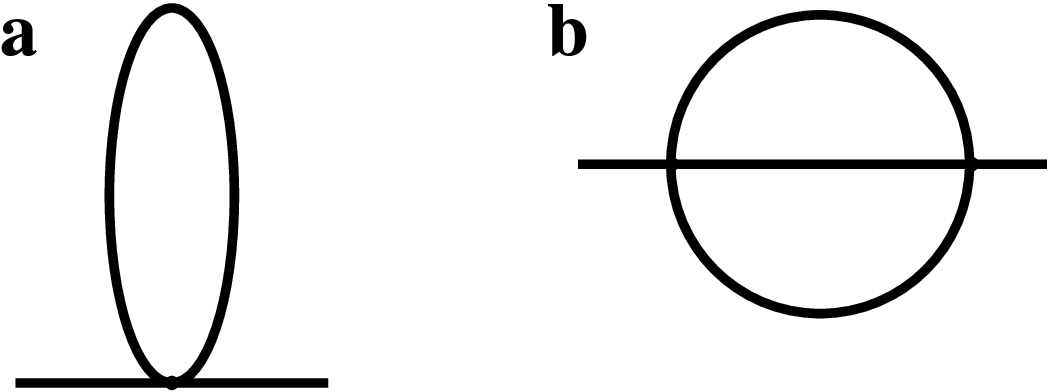}}}
\vspace*{1cm}
\caption{Diagrams representing low-order terms
in a perturbation series for the dynamics generated by
Eq.~2.  Lines represent instantaneous correlations
of the field at two points in space, 
while vertices represent interactions.}
\label{fig:diagrams}
\end{figure}

Treating $\lambda$ as a perturbation parameter, the solution to
Eq.~\ref{equ:langevin} may be written as an infinite series of terms,
each representing a collection of field interactions and periods of
free propagation.  As a result, $C_{\bf k}(t)$ and its associated
response function, $G_{\bf k}(t) = -\beta\, d C_{\bf k}(t)/dt$ can be
expanded in powers of $\lambda$.  (We focus exclusively on the portion
of the phase diagram in which dynamics are ergodic, so that $C_{\bf
k}(t)$ and $G_{\bf k}(t)$ are related by the fluctuation-dissipation
theorem.)  Fig.~\ref{fig:diagrams} shows diagrammatic representations
of the first two terms in the series for $C_{\bf k}(t)$.  The
development of this expansion, as well as the partial series
resummations underlying our approximations, have been discussed
thoroughly in the context of other models.  We will describe
physically significant highlights of the procedure and refer the
reader to Refs.\cite{grousson_etal2,bouchaud} for details.

A linear equation of motion for $C_{\bf k}(t)$ results from summing
only terms in the series whose diagrams have the basic topology shown
in Fig.~\ref{fig:diagrams}(a).  In the irreducible segments of these
``tadpole'' diagrams, all interactions coincide in time.
Consequently, such a summation renormalizes only the static portion of
the basic tadpole diagram of Fig.~\ref{fig:diagrams}(a).  The dynamics
predicted by this Hartree resummation scheme are identically those of
a variationally optimized harmonic reference system\cite{deem}.  They
are thus obtained more directly by assuming Gaussian statistics for
$\phi_{\bf k}(t)$.  Specifically, we multiply the Fourier transform of
Eq.~\ref{equ:langevin} by $\phi_{-{\bf k}}(0)$ and average over the
noise history, yielding
\begin{eqnarray}
&& \hspace{-10mm}{\partial C_{\bf k}(t)\over \partial t} = -[\tau +
k_0^{-2}(k^2-k_0^2)^2] C_{\bf k}(t) + 
\nonumber \\
&&
{\lambda \over 3!} \sum_{{\bf
k}',{\bf k}''} \langle \phi_{{\bf k}'}(t) \phi_{{\bf k}'}(t)
\phi_{{\bf k}-{\bf k}'-{\bf k}''}(t) \phi_{-{\bf k}}(0) \rangle .
\label{equ:avlang}
\end{eqnarray}
Eq.~\ref{equ:avlang} is the first member of a complicated hierarchy of
equations relating multi-point fluctuations to correlations of higher
order.  But if we assume that $\phi_{\bf k}(t)$ is a Gaussian random
variable, the hierarchy closes immediately:
\begin{equation}
{\partial C^{\rm H}_{{\bf k}}(t) \over \partial t} = -\mu_{\bf k}^{\rm
H}C^{\rm H}_{{\bf k}}(t).
\label{equ:hartree}
\end{equation}
Here, the renormalized mass $\tau_{\rm H}$ that appears in the
expression for the structure factor $\mu_{\bf k}^{\rm H} = 1/C^{\rm
H}_{\bf k}(0) = \tau_{\rm H} +k_0^{-2}(k^2-k_0^2)^2$ is determined
self-consistently by
\begin{equation}
\tau_{\rm H}=\tau +{\lambda \over 2} \sum_{{\bf k}'} C^{\rm H}_{{\bf
k}'}(0).
\label{equ:braz}
\end{equation}
This result is precisely Brazovskii's static
approximation\cite{brazovskii}.  The relaxation described by
Eq.~\ref{equ:hartree}, while simply exponential, occurs with rates
that are significantly renormalized by the entropy of fluctuations
near $|{\bf k}|= k_0$.

More elaborate, nonlinear approximations for $C_{\bf k}(t)$ result
from incorporating diagrams with more complicated
topologies\cite{grousson_etal2,bouchaud}.  The mode coupling
approximation (MCA) is an example, including diagrams with the
``sunset'' shape of Fig.~\ref{fig:diagrams}(b).  Summing all terms
that renormalize the propagators (but not the vertices) of the basic
sunset diagram yields the MCA.  Because these contributions involve
more than one unique time variable, they are capable in principle of
capturing nontrivial memory effects.  As shown in Ref.\cite{bouchaud},
the self-consistent result of this resummation is
\begin{eqnarray}
&&
\hspace{-5mm}
\frac{\partial C^{\rm MCA}(t)_{{\bf k}}}{\partial t}=  -\mu^{\rm
MCA}_{{\bf k}} C^{\rm MCA}_{{\bf k}}(t) + \nonumber \\ & &
\frac{\lambda^{2}\beta}{6}\int_{0}^{t}dt^{'}\sum_{{\bf k}',{\bf k}''}
\left[C^{\rm MCA}_{{\bf k}'}(t-t') C^{\rm MCA}_{{\bf k}''}(t-t')
\right.
\nonumber \\
&& \left. \times
C^{\rm MCA}_{{\bf k}-{\bf k}'-{\bf k}''}(t-t')\right] \frac{\partial
C^{\rm MCA}_{\bf k}(t')}{\partial t^{'}}.
\label{equ:mca}
\end{eqnarray}
The final, nonlinear term of Eq.~\ref{equ:mca} explicitly couples the
dynamics of fluctuations at different wave vectors, so that the decay
of $C^{\rm MCA}_{\bf k}(t)$ is not simply exponential.  The
mode-coupling estimate of the static structure factor, $C^{\rm
MCA}_{\bf k}(0) = 1/\mu^{\rm MCA}_{\bf k}$, is determined by a
self-consistent equation involving both the ``tadpole'' and ``sunset''
diagrams.  We avoid this static calculation by instead replacing
$\mu^{\rm MCA}_{\bf k}$ in Eq.~\ref{equ:mca} with the exact form of
$C_{\bf k}^{-1}(0)$ from numerical simulations.  This procedure is
commonly employed in mode-coupling studies of supercooled liquids.

As the microphase transition point is approached from high
temperature, we expect that only modes near the ordering wave vector
$k_{0}$ will remain important.  In this regime, a reduced model
without reference to the coupling of specific length scales should
capture the qualitative behavior of Eq.~\ref{equ:mca}.  Such a
schematic model is similar to that studied by Leutheusser for
structural glass-forming liquids\cite{leutheusser}.  (Indeed,
Eq.~\ref{equ:mca} is only slightly different from that encountered in
the idealized mode-coupling theory of supercooled
liquids\cite{goetze}.  In particular, the memory kernel involves a
two-point correlation function raised to the third, rather than
second, power.)  Restricting attention to $|{\bf k}|=k_0$, and
neglecting coupling to other wave vectors, Eq.~\ref{equ:mca} reduces
to the dynamical equation exactly describing the $p$-spin model of a
mean field spin glass (with $p=4$).  Since for $p > 2$ the critical
properties of such models are essentially $p$-independent, we expect
that near a critical point, Eq.~\ref{equ:mca} will exhibit a plateau
and an eventual transition to nonergodic behavior\cite{young}.

The character of the slow dynamics resulting from the theories
underlying Eq.~\ref{equ:hartree} and Eq.~\ref{equ:mca} are
fundamentally different.  The dynamical Hartree theory
(Eq.~\ref{equ:hartree}) may exhibit a rapid slowing of dynamics as a
function of inverse temperature {\em only} if the statics, as
expressed through the renormalized mass $\tau_{\rm H}$, are strongly
temperature dependent.  On the other hand, due to the nonlinearity of
Eq.~\ref{equ:mca}, a slight change in the structure factor may result
in a dramatic change in relaxation times.  It is well known that
glass-forming liquids show little change in static structure as the
glass transition is approached\cite{nagel}.  Thus, theories of the
type given in Eq.~\ref{equ:hartree} are not relevant near the glass
transition.  In the following sections, the predictions of
Eq.~\ref{equ:hartree} and Eq.~\ref{equ:mca} will be compared with
simulations for the Coulomb-frustrated system.

The renormalized perturbation theories developed in this section are
strictly valid only in the limit of weak coupling, i.e., for small
$\lambda$ or large, positive $\tau$.  In our simulation work, we fix
$\lambda=1$.  It is thus instructive to ask at what value of $\tau$
these theories are expected to break down.  To answer this question,
we follow the arguments of Hohenberg and Swift\cite{hohenberg}.
Specifically, we compare the Hartree approximation to the renormalized
mass, $\tau_{\rm H}$, with corrections introduced by mode coupling
(i.e., the renormalized sunset diagram).  These corrections are
comparatively small when
\begin{equation}
|\tau| \lesssim 0.2 \, k_0^{7/5}
.
\label{equ:breakdown}
\end{equation}
In this regime the Hartree and mode coupling approximations 
are controlled, and differ only quantitatively from one another.  
For larger $|\tau|$, however, the two
approximations can differ substantially, as we will see in numerical
results presented in the following section.  It has been noted
previously that the static Hartree approximation can be accurate
beyond its strict range of validity.  There is thus no guarantee that
a range of $\tau$ exists in which the mode coupling approximation
significantly improves upon an appropriately chosen harmonic reference
system.

\section{Simulations}
In order to follow the dynamics of Eq.~\ref{equ:langevin} or
Eq.~\ref{equ:mc} numerically, it is necessary first to 
coarse-grain the field $\phi({\bf r},t)$ in space.  This procedure
yields a (periodically replicated) finite set of dynamical
variables, whose time evolution may be integrated approximately
over short intervals.  We select a coarse-graining length
$a = 2\pi/n k_0$, and define new fields at lattice points
${\bf r}_i$:
\begin{equation}
\Phi_i(t) = a^{-3} \int_{v_i} d{\bf r} \, \phi({\bf r},t),
\label{equ:coarse}
\end{equation} 
where $v_i$ is bounded by a cube of side length $a$ centered at ${\bf
r}_i$.  In the calculations described below, $n=8$, so that a domain
of wavelength $\pi/k_0$ comprises several ``soft spins'' $\Phi_i$.  To
lowest order in $a$ and a small time increment $\Delta t$, these
renormalized fields evolve according to
\begin{eqnarray}
\Phi_i(t+\Delta t) = && \Phi_i(t) - \Delta t[(\tau + k_0^{-2}({\cal L}+
k_0)^2) \Phi_i(t)  
\nonumber \\
&&
+ {\lambda\over{3!}}\Phi_i^3(t)]
+\overline{\eta}_i(t).
\label{equ:discrete}
\end{eqnarray}
Here, the lattice approximation to the Laplacian operator acting on a
function of space, ${\cal L}f(x_i) = a^{-2}\sum_{j \in {\rm nn}}
f(x_j) - f(x_i)$, is taken to include a sum over nearest neighbors
only (denoted nn).  After coarse-graining, statistics of the random
force remain Gaussian, with $\langle \overline{\eta}({\bf r}_i,t_m)
\overline{\eta}({\bf r}_j,t_n) \rangle = 2 \beta^{-1} (\Delta t/
a^3)\delta_{ij} \delta_{mn}$.

The simulation algorithm described above (which is very similar to
those of Refs.\cite{chakraborty,elder}) has several advantages over
the numerical approach of Ref.\cite{grousson_etal}, which employs
``hard spins'' ($\Phi_i=\pm 1)$ and explicit frustration.  First,
spins interact only with nearest and next-nearest neighbors, providing
linear scaling of computational effort with system size.  Because
cumbersome techniques associated with long-ranged forces are not
required, a larger set of dynamical variables may be considered.  In
our calculations, the periodically replicated unit cell includes
$64^3$ spins arranged on a cubic lattice.  More importantly, the
coarse-graining procedure allows the dimensions of the unit cell to
scale with the physically relevant length $k_0^{-1}$.  As a result,
the unit cell spans several correlation lengths, even for very small
values of $k_0$.  By contrast, in the work of
Ref.\cite{grousson_etal}, the unit cell is comparable to a single
natural lamellar spacing for several of the simulated states
(particularly those corresponding to ``fragile'' systems).  In those
cases, significant finite size effects are possible.  In effect,
Grousson {\em et al.}  cut off slowly relaxing fluctuations at small
${\bf k}$ rather than the rapidly relaxing fluctuations at large ${\bf
k}$ that are integrated out in our approach.  In Sec. 5 we discuss the
dynamical implications of such a cutoff.

\begin{figure}
\centerline{\resizebox{8cm}{!}{\includegraphics{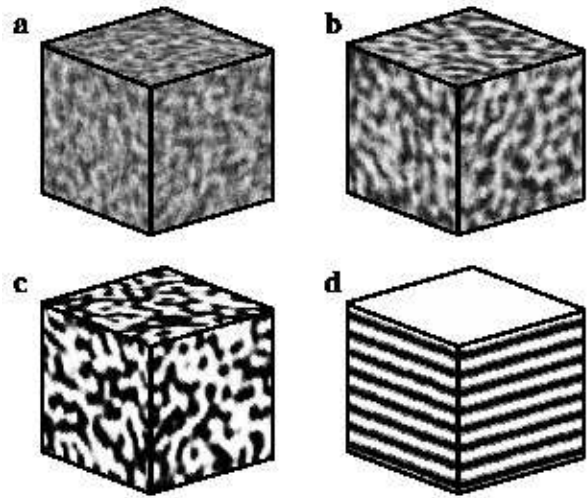}}}
\caption{Representative equilibrium configurations 
of the model system 
for $k_0$=0.5 and $\tau=0$ ({\bf a}), 
$\tau=-0.12$ ({\bf b}), and
$\tau< \tau_{\rm tr}$ ({\bf d}).
The configuration depicted in ({\bf c})
was obtained from the nonequilibrium evolution
of a nearly random state quenched
instantaneously to $\tau< \tau_{\rm tr}$.}
\label{fig:pics}
\end{figure}

Using Eq.~\ref{equ:discrete}, we have computed the dynamics of several
states of the model system at temperatures above the microphase
separation transition ($\tau \gtrsim \tau_{\rm tr}$).  We focus on
three values of $k_0$ (0.1, 0.5 and 1.0) corresponding to a somewhat
broader range of model parameters than was considered in
Ref.\cite{grousson_etal}.  In each case, $\lambda=1$, so that the
microscopic dynamics is in principle strongly nonlinear.
Representative configurations of the system are depicted in
Fig.~\ref{fig:pics}, typifying the high-temperature paramagnetic phase
(a), the disordered phase near the microphase separation transition
(b), the ordered lamellar phase (c), and a nonequilibrium state
produced by rapid quenching of a disordered system to low temperature
(d).

\begin{figure}
\centerline{\resizebox{7cm}{!}{\includegraphics{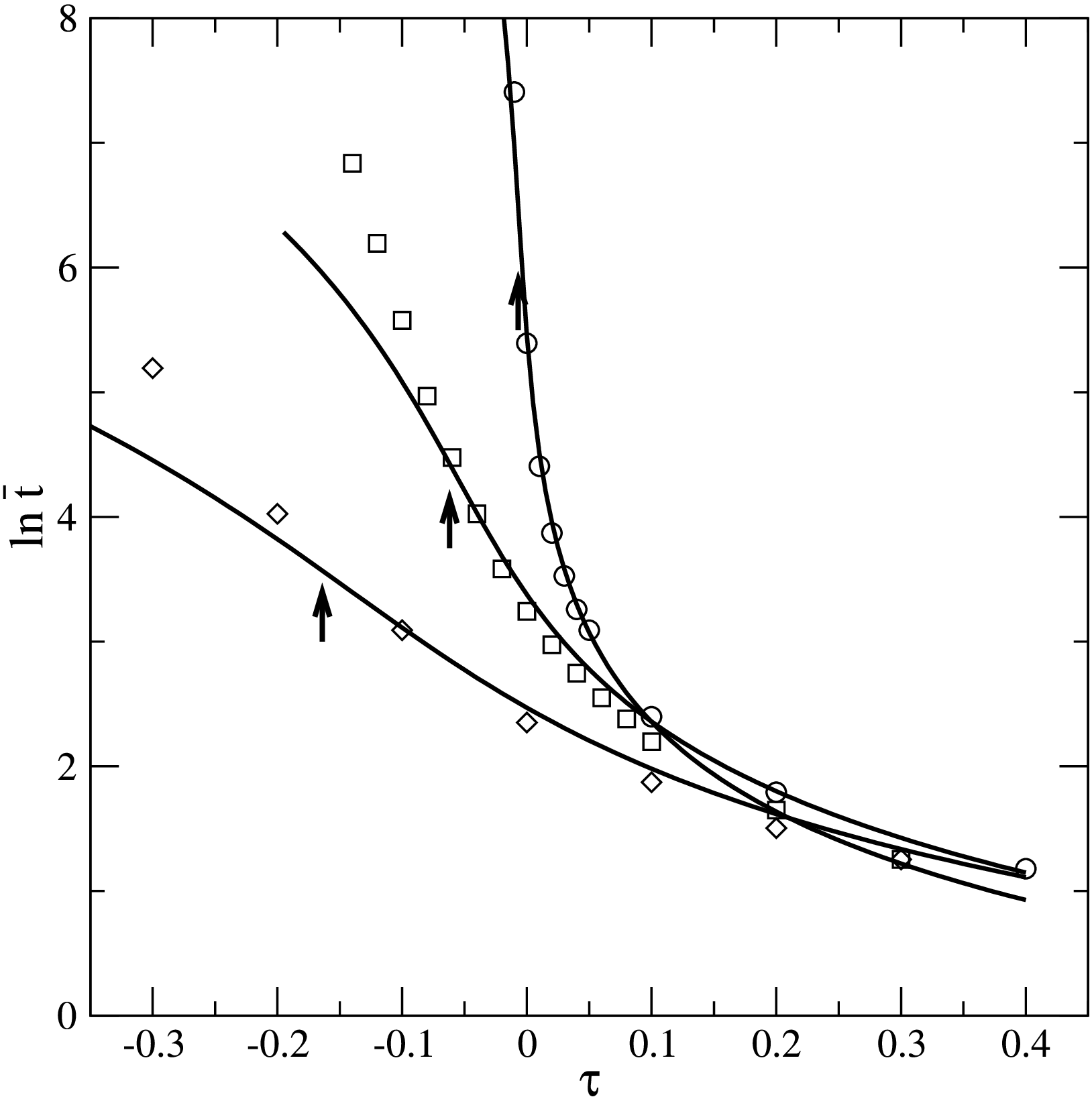}}}
\caption{Time $\overline{t}$ 
required for spin correlations 
to decay to 10\% of their initial values, as a
function of scaled temperature $\tau$.
Circles, squares, and diamonds show numerical results
for $k_0=0.1$, $k_0=0.5$, and $k_0=1.0$, respectively.
Solid lines are predictions of the Hartree approximation
described in the text.  Arrows indicate temperatures
at which this approximation is expected to break down
for each value of $k_0$,
as estimated using Eq.~\ref{equ:breakdown}.}
\label{fig:relax}
\end{figure}

For each value of $k_0$ we consider, the relaxation of spin
correlations slows dramatically near the transition to microphase
separation.  The time required for single-spin correlation,
\begin{equation}
C(t) \equiv \langle \Phi_i(0) \Phi_i(t) \rangle = N^{-1} \int \frac{d
 {\bf k}}{8\pi^{3}}C_{\bf k}(t),
\label{equ:ct}
\end{equation}
to decay to $10\%$ of its initial value, $\overline{t}$, is plotted as
a function of $\tau$ in Fig.~\ref{fig:relax}.  The growth of
relaxation times as $\tau$ approaches $\tau_{\rm tr}$ is sharpest for
the smallest value of $k_0$.  Indeed, critical fluctuations are
suppressed least strongly in this case, as evidenced by the onset of
sluggishness very near $\tau=0$.

Grousson {\em et al.} have likened systems corresponding to large and
small values of $k_0$ to ``strong'' and ``fragile'' glass formers,
respectively\cite{grousson_etal}.  For fragile cases, they have even
shown that the temperature dependence of $\overline{t}$ is well fit by
the Vogel-Fulcher form found for supercooled liquids.  Although this
functional form suggests an emerging importance of activated
processes, the dramatic growth of relaxation times is in fact well
captured by the harmonic reference system described in Sec.~3.
Numerical solutions of Eq.~\ref{equ:braz} (plotted as solid lines in
Fig.~\ref{fig:relax}), corresponding to this Hartree approximation,
are especially accurate in the most fragile case ($k_0=0.1$).  Even
for the least fragile case ($k_0=1$), computed rates differ from
predicted values by at most a factor of two.  Activated barrier
crossing is manifestly absent on a harmonic landscape, strongly
implying that slow dynamics are driven by static renormalization,
rather than by fundamental changes in the structure of trajectory
space\cite{chandler}.  The static structure factor, $C_{\bf k}(0)$, in
fact becomes more sharply peaked in a way that mirrors the sudden
growth in $\overline{t}$.  In other words, the slowing of relaxation
appears to be an example of de~Gennes narrowing\cite{hansen}.

\begin{figure}
\centerline{\resizebox{7cm}{!}{\includegraphics{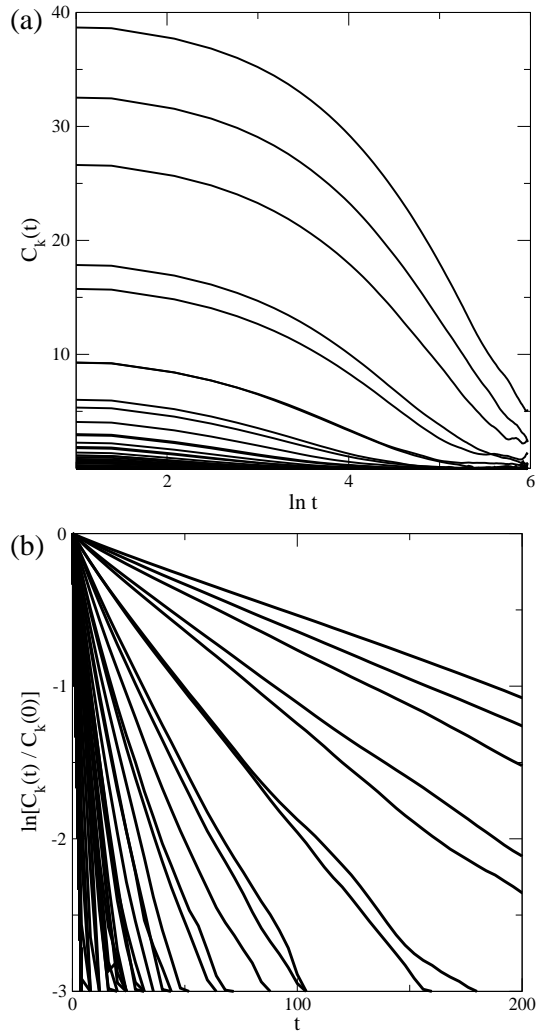}}}
\caption{$C_{\bf k}(t)$ and 
$\ln{[C_{\bf k}(t)/C_{\bf k}(0)]}$ for
several $|{\bf k}|$ as a function of $\ln{t}$ 
and $t$, respectively, for
the state $k_0=0.5$, $\tau=-0.1$.  The uppermost 
curve in each panel
corresponds to $|{\bf k}|=k_0$ and exhibits the slowest decay.}
\label{fig:ckt}
\end{figure}

The time dependence of spin correlations provides further evidence for
this interpretation.  Specifically, even when relaxation is very slow,
correlations decay nearly exponentially at each wave vector.  In
Fig.~\ref{fig:ckt}, $C_{\bf k}(t)$ is plotted for many $k$ values for
a system very near microphase separation ($k_0=0.5$, $\tau=-1$).
Included wave vectors span a range from the lowest accessible spatial
frequency ($k=2\pi/L$) to several multiples of $k_0$.  In no case is
relaxation detectably caged or stretched at long times.  The
single-spin correlation function, $C(t)$, in Eq.~\ref{equ:ct} is a
superposition of all $C_{\bf k}(t)$, and thus does not decay as a
single exponential.  At long times, however, relaxation is dominated
by the slowest modes, those with wave vectors lying in a spherical
shell with $|{\bf k}|\approx k_0$, and is very nearly exponential.
Since static correlations are strongest for these modes, especially
near the microphase separation transition, $C(t)$ is nonexponential
only over a small range of the total decay.

Remarkably, spin relaxation is essentially identical for a very
different choice of microscopic propagation rules which conserve the
field $\Phi$.  In this Monte Carlo dynamics, described in Sec.~2, each
trial move simultaneously displaces $\Phi$ at a randomly chosen site
$i$ and at a a site $j$ randomly chosen from the nearest neighbors of
$i$.  The displacement at $i$, $\Delta \Phi_i$, is exactly compensated
by that at $j$, i.e., $\Delta \Phi_j=-\Delta \Phi_i$.  In this way,
$\Phi$ is conserved at all length scales greater than or equal to the
lattice spacing $a$.  In general, such a constraint can influence
dynamical behavior dramatically.  For instance, scaling exponents for
unstable domain growth in similar models depend intimately on the
conservation of order parameters\cite{bray}.  The relaxation described
above, however, is modified by the constraint only at very short
times.

The insensitivity of slow dynamics to field conservation in this model
was anticipated by Sachdev, who noted that the corresponding
constraint couples strongly only to fluctuations with very small wave
vector ($|{\bf k}|\approx 0$)\cite{sachdev}.  The long-lived
correlations in Fig.~\ref{fig:ckt}, however, are governed not by these
modes, but instead by fluctuations of finite wave vector ($|{\bf
k}|\approx k_0$), which couple relatively weakly to the constraint.
Conservation is thus of only modest importance at long times in the
{\em disordered} phase.  These facts can lead in principle to dramatic
finite size effects in numerical simulations.  Specifically, the
smallest periodically replicated unit must accommodate fluctuations
with wavelengths several times $k_0$.  Otherwise, the absence of truly
long wavelength fluctuations will generate spuriously strong coupling
of conservation constraints to the slowest accessible modes.  As a
result, artificial dynamical features may appear at long times.  For
model energetics that are similar to but different from
Eq.~\ref{equ:hamiltonian}, stretched exponential relaxation has been
computed from simulations in which system dimensions are comparable to
$k_0^{-1}$\cite{grousson_etal}.  To the extent that fluctuations near
$|{\bf k}|=k_0$ are independent of model details, such anomalous
dynamics should not survive in the thermodynamic limit.

\begin{figure}
\centerline{\resizebox{7cm}{!}{\includegraphics{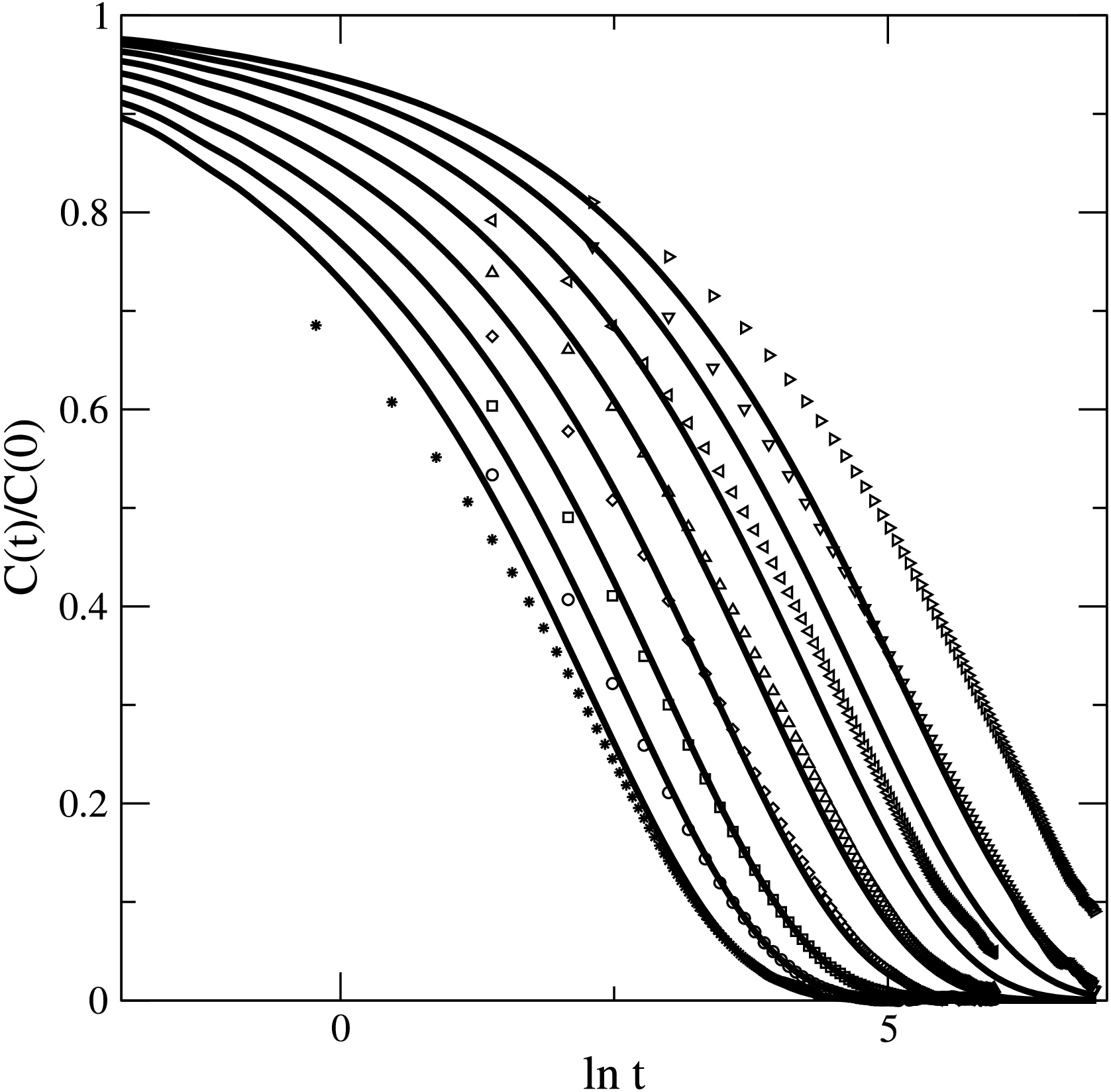}}}
\caption{Decay of spin correlation $C(t)$
predicted by Hartree approximation (lines), compared
with results of numerical simulations (symbols) for
several thermodynamic states:
$\tau=-0.14$, $\tau=-0.12$, $\tau=-0.1$, $\tau=-0.08$, 
$\tau=-0.06$, $\tau=-0.04$, $\tau=-0.02$, $\tau=0$, 
(in order from top to bottom in the plot).
In each case, $k_0=0.5$.}
\label{fig:hartree}
\end{figure}

The persistence of exponential relaxation into the neighborhood of
microphase separation, obtained numerically both for Langevin and
Monte Carlo dynamics, is consistent with the dynamical Hartree
approximation described in Sec. 3.  In fact, this simple theory also
predicts with remarkable accuracy the time scales of relaxation, even
as they grow by several orders of magnitude.  Results for the case
$k_0 = 0.5$ are compared in Fig.~\ref{fig:hartree} by plotting
$C(t)/C(0)$ obtained from theory and simulation for several values of
$\tau$.

The success of the harmonic reference system provides further evidence
that ``vitrification'' in this model is driven by dramatic changes in
structural order rather than novel relaxation mechanisms.  Indeed,
several dynamical features that would seem to be related to activated
barrier crossing may be well rationalized in the Hartree picture.  For
example, the fragility parameter defined empirically by
\begin{equation}
\overline{t} \propto \exp\left({T_{\rm K}D \over T-T_{\rm K} }\right)
\label{equ:fragility}
\end{equation}
appears to scale as $D\sim Q^{1/2}$ in the simulations of
Ref.\cite{grousson_etal}.  Here, $T_{\rm K}$ is a fitted Kauzmann
temperature at which relaxation times appear to diverge. Schmalian and
Wolynes have suggested that this scaling arises from the entropy of
activation for structural rearrangement of mesoscopic
domains\cite{schmalian,schmalian2,schmalian3}.  The Hartree
approximation offers a simpler explanation.  Fitted relaxation times
follow $\overline{t} \sim \tau_{H}^{-1} \sim \exp(D/\tau)$ rather
well.  It is easy to show by dimensional scaling that the renormalized
mass is a function only of $\tau / Q^{1/3}$, yielding immediately $D
\sim Q^{1/3}$\cite{scaling_footnote}.  This slightly different scaling
form fits the results of Grousson {\em et al.} equally
well\cite{grousson_etal}.  For the simulations of larger systems we
have presented, the Hartree prediction appears to be superior.

\begin{figure}
\centerline{\resizebox{7cm}{!}{\includegraphics{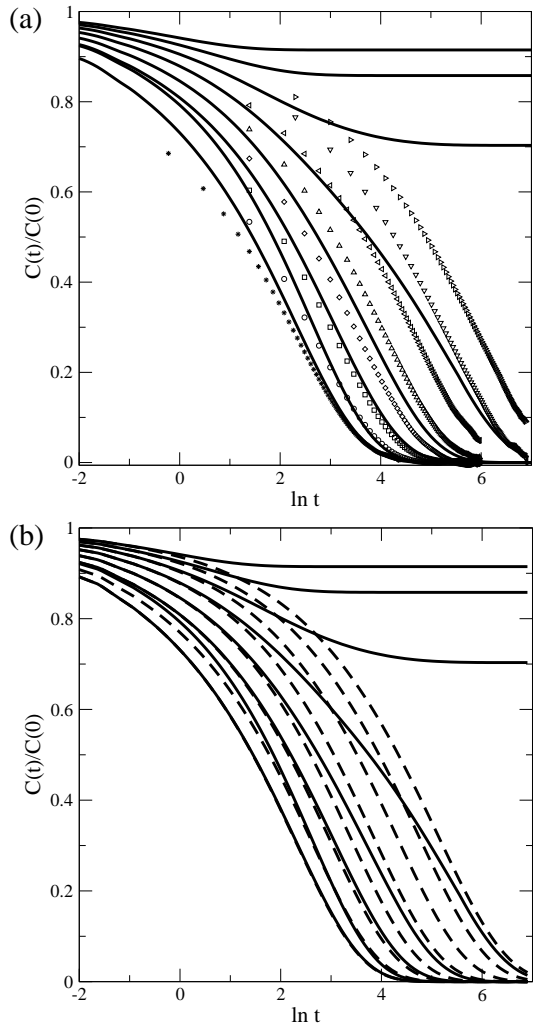}}}
\caption{Comparison of MCA (solid lines) with numerically simulated
dynamics (symbols) (a) and MCA (solid lines) with Hartree
approximation (dashed lines) (b).  Results are shown for thermodynamic
states identical to those plotted in Fig.~5.  According to
Eq.~\ref{equ:breakdown}, contributions of mode coupling become
significant around $\tau \simeq -0.06$ (fifth line from the top).
Note that this temperature is very near the critical temperature at
which MCA predicts loss of ergodicity.}
\label{fig:mca}
\end{figure}

The dynamical scenario predicted by the simplified mode coupling
theory of Sec. 3, on the other hand, is not borne out in our
simulations.  Most significantly, we observe neither loss of
ergodicity nor two-step relaxation over the relevant range of $\tau >
\tau_{\rm tr}$.  These failures of the dynamically nonlinear
approximation are evident in Fig.~\ref{fig:mca}, comparing the MCA and
simulation results for the same system and thermodynamic states
considered in Fig.~\ref{fig:hartree}.  Interestingly, the MCA predicts
trapping at values of $\tau$ for which the Hartree approximation
remains reasonable.  The infinite series of terms incorporated in the
mode coupling approximation thus adds little realism to the lower
order description, and eventually leads to incorrectly anomalous
behavior.  This series of terms must be compensated to a large degree
by omitted terms at each order.  There have been suggestions that
similar cancellation occurs in mode coupling expansions of supercooled
liquid dynamics\cite{oppenheim}.  In that case, however, signatures of
idealized mode coupling (i.e., two-step relaxation) survive despite
the existence of omitted relaxation channels.  In our model, no such
signatures are evident.

We emphasize that the thermodynamic states we have simulated lie
exclusively in the disordered phase of our model system.  We have
confirmed this fact by computing the work to reversibly impose
long-range order (i.e., nonzero $\langle |\Phi_{\bf k}| \rangle$ at
$|{\bf k}|=k_0$).  Although the static Hartree approximation suggests
that the states of lowest temperature considered for each $k_0$ have
global free energy minima in ordered configurations, the computed free
energy of the paramagnetic state is in fact lower for each case.  This
quantitative failure of Brazovskii's approximation is not surprising,
as the relevant states lie outside the strict range of validity of the
approximation for $\lambda=1$\cite{brazovskii,hohenberg}.  Our
analysis thus leaves unexplored a narrow window of the disordered
state extremely close to $\tau = \tau_{\rm tr}$.  While it is possible
that qualitatively new dynamical behavior arises in this region, it
cannot account for the dramatic slowing down we have demonstrated at
higher temperature, which is driven by extreme structural changes
rather than activated processes.  Furthermore, the sudden onset of
nontrivial behavior would be in stark contrast to the more gradual
onset of caging and stretched exponential behavior in real liquids.

\section{Conclusions}
We have examined in detail a simple model that is closely related to
the frustration-limited domain theory of Kivelson, Tarjus and
coworkers\cite{fldt}, and to the uniformly frustrated ``stripe-glass''
model studied by Schmalian and
Wolynes\cite{schmalian,schmalian2,schmalian3}.  The disordered phase
of this model system indeed displays some hallmarks of molecular
glass-forming liquids.  But its dynamics differ qualitatively from
generic glassy behavior in several respects.

While relaxation times increase dramatically in these models in a
non-Arrhenius fashion as temperature is lowered, we find that an
optimized harmonic reference system captures the time dependence of
fluctuations semi-quantitatively.  In contrast to the vitrification
of molecular liquids, the onset of this sluggishness is {\em not}
accompanied by significant power law or stretched exponential decay of
correlations in time.  Perhaps most importantly, we find that the the
slow decay of dynamical correlations is driven by significant changes
in static structure.  An idealized mode-coupling theory captures these
changes less accurately than the simpler harmonic approach, predicting
caging and eventual trapping at temperatures where the simulated
dynamics remain exponential and ergodic.  Together, these results
strongly suggest that slow dynamics of this model system arise from
the same Gaussian fluctuations that drive the microphase separation
transition first order.  This conclusion differs markedly from those
of Kivelson {\em et al.}\cite{fldt}, Grousson {\em et 
al.}\cite{grousson_etal,grousson_etal2}, and of Schmalian and
Wolynes\cite{schmalian,schmalian2,schmalian3}, which invoke activated
barrier crossing to explain the dramatic temperature dependence of
relaxation times.  Uniform frustration alone thus appears insufficient
to account for the unusual relaxation properties of supercooled
liquids.  It remains possible, of course, that such frustration plays
a key role, but in concert with other important physical mechanisms
and constraints.

The simulations described in this paper constitute a thorough study of
the thermodynamically disordered phase of a uniformly frustrated
model.  They raise some interesting questions about such models and
leave others unaddressed.  In particular, while glassy behavior is not
manifest for $\tau \gtrsim \tau_{\rm tr}$, we can not rule out the
existence of caging or trapping extremely close to or below the
microphase separation temperature (since relaxation at these
temperatures is prohibitively slow).  To explore this possibility, we
have simulated several initially disordered states with $\tau <
\tau_{\rm tr}$.  Their nonequilibrium evolution towards long-ranged
order exhibits the self-similar aging characteristic of coarsening
phenomena, but for numerically accessible time scales does not
resemble the relaxation of a supercooled liquid.  Nonexponential
equilibrium relaxation could yet emerge at temperatures above but {\em
very} near the order-disorder transition\cite{schmalian4}.  Eastwood
and Wolynes have in fact recently suggested that, for certain spin
models, surface tension effects drive the onset of activated dynamics
nearly to the ideal glass transition
temperature\cite{eastwood_wolynes}.  Although quite interesting, such
behavior would itself be uncharacteristic of glass-forming liquids, as
it would appear only within an extremely narrow thermodynamic range.

\section*{Acknowledgments}
We would like to thank B. Chakraborty, G. Tarjus and P. Wolynes for
useful discussions.  D.R.R. was supported by an NSF CAREER Award
(\#0134969). P.L.G. is an M.I.T. Science Fellow.

\bibliographystyle{prsty}

\end{document}